\documentclass[floatfix,aps,twocolumn,superscriptaddress,preprintnumbers,amsmath,amssymb]{revtex4}

\usepackage[applemac]{inputenc}
\usepackage[T1]{fontenc}
\usepackage{amsfonts, amssymb, amsmath, dsfont}
\usepackage{graphicx}
\usepackage{float}
\usepackage{ulem}
\usepackage{color}

\newcommand{\bPhi}{\mathbf{\Phi}}

\newcommand{\bE}{\mathbf{E}}

\newcommand{\bG}{\mathbf{G}}

\newcommand{\bv}{\mathbf{v}}

\newcommand{\br}{\mathbf{r}}

\newcommand{\bp}{\mathbf{p}}

\newcommand{\bu}{\mathbf{u}}

\newcommand{\be}{\mathbf{e}}

\def\d{\mathrm{d}}
\def\eps{\varepsilon}
\def\Re{\mathrm{Re}}
\def\Im{\mathrm{Im}}

\renewcommand{\emph}{\textit}

\begin{document}

\title{Cross density of states and mode connectivity: Probing wave localization \\ in complex media}

\author{Antoine Canaguier-Durand}
\affiliation{Laboratoire Kastler Brossel, Sorbonne Universit\'e, CNRS, ENS-PSL University, Coll\`ege de France, Paris, France.}
\author{Romain Pierrat}
\affiliation{ESPCI Paris, PSL University, CNRS, Institut Langevin, 1 rue Jussieu, F-75005, Paris, France.}
\author{R\'emi Carminati}
\altaffiliation{remi.carminati@espci.fr}
\affiliation{ESPCI Paris, PSL University, CNRS, Institut Langevin, 1 rue Jussieu, F-75005, Paris, France.}

\begin{abstract} 
   We introduce the mode connectivity as a measure of the number of eigenmodes of a wave equation connecting two points
   at a given frequency. Based on numerical simulations of scattering of electromagnetic waves in disordered media, we
   show that the connectivity discriminates between the diffusive and the Anderson localized regimes. For practical
   measurements, the connectivity is encoded in the second-order coherence function characterizing the intensity emitted
   by two incoherent classical or quantum dipole sources. The analysis applies to all processes in which spatially
   localized modes build up, and to all kinds of waves.
\end{abstract}

\maketitle

\section{Introduction}

Spatially localized modes are key elements in the description of many phenomena in mesoscopic and wave
physics~\cite{Sheng-book}, and their control is a central issue in photonics, acoustics or microwave engineering.
Indeed, wave transport through disordered media is substantially affected by Anderson localization~\cite{Localization}.
Surface-plasmon modes on percolated metallic films also undergo a localization process in certain conditions, producing
a subwavelength concentration of energy of interest in nanophotonics~\cite{Shalaev2000}. In photonics and acoustics
(phononics), bandgaps in periodic structures~\cite{bandgaps, Vos2015} and cavity modes are used to enhance wave-matter
interaction, e.g., to enter the regimes of cavity quantum electrodynamics (QED)~\cite{CavityQED} or optomechanics
\cite{Arcizet2009,Aspelmayer2014}. An underlying question in the description of localization phenomena, and in the
design of artificial structures producing localized modes, is the characterization of an isolated mode (e.g., as a
signature of Anderson localization itself, or to reach cavity QED regimes~\cite{strong-coupling,Caze2013-sc}), and the
measure of the spatial connection between individual modes (e.g., in the description of transport through a chain of
weakly connected modes~\cite{necklace-states}).

In this article, we introduce the concept of mode connectivity, defined from the local and cross densities of states
(LDOS and CDOS), as a measure of the connection between two points sustained by the eigenmodes at a given frequency. We
focus on electromagnetic waves, but the concept applies to other kinds of waves. As an illustration, we show that the
mode connectivity allows one to discriminate between diffusive transport and Anderson localization in disordered media.
We then define observables that depend directly on the connectivity and the LDOS, and that could be measured from the power
emitted by two dipole sources placed inside the medium or at close proximity to its surface. We examine separately the
case of classical sources (antennas) and of quantum emitters (single-photon sources). The proposed approach could
provide an unambiguous probe of Anderson localization of electromagnetic waves in 3D, whose existence remains a debated
issue~\cite{Skipetrov2014}.

\section{Cross density of states and connectivity}

For a monochromatic electromagnetic field at a frequency $\omega$, we define the CDOS $\rho_{12}$ as~\cite{Caze2013-CDOS}
\begin{align}\label{def_CDOS}
   \rho_{12}= \frac{2 \omega}{\pi c^2} \Im \left [\bu_1 \cdot \bG(\br_1,\br_2,\omega) \bu_2 \right ]
\end{align}
where $\br_1$ and $\br_2$ are two points, $\bu_1$ and $\bu_2$ are two unit vectors, and $c$ is the speed of light in
vacuum. In this expression, $\bG(\br_1,\br_2,\omega)$ is the electric Green function, that connects an electric-dipole
source $\bp$ at point $\br_2$ to the electric field generated at point $\br_1$ through the relation $\bE(\br_1) = \mu_0
\, \omega^2 \, \bG(\br_1,\br_2,\omega) \bp$. Note that we consider here a partial CDOS, projected over two polarization
directions $\bu_1$ and $\bu_2$, that differs from the full CDOS initially introduced in Ref.~\cite{Caze2013-CDOS} in
which the polarization degrees of freedom are averaged out.

The CDOS measures the connection between point $\br_1$ and $\br_2$ sustained by the
eigenmodes~\cite{Caze2013-CDOS,Sauvan2014}. Interestingly, the CDOS obeys the inequality
\begin{align}\label{CS_inequality}
   \rho_{12} \leq \sqrt{ \rho_{11} \rho_{22}}
\end{align}
where $\rho_{ii} = 2 \omega/(\pi c^2) \Im \left [\bu_i \cdot \bG(\br_i,\br_i,\omega) \bu_i \right ]$ is the (partial)
LDOS that measures the weighted contribution of eigenmodes at position $\br_i$, projected along the polarization
direction $\bu_i$~\cite{Carminati-review}.  The inequality (\ref{CS_inequality}) was initially derived in
Ref.~\cite{Canaguier2016}, and the proof is recalled in Appendix~\ref{cdos_inequality}, for consistency.  This
naturally leads to the introduction of the mode connectivity
\begin{align}\label{def_connectivity}
   {\cal C}_{12} = \frac{\left| \rho_{12}  \right|}{ \sqrt{\rho_{11} \rho_{22} }} ,
\end{align}
a dimensionless number lying within $[0,1]$. The two observation points $\br_1$ and $\br_2$ are highly connected at
frequency $\omega$ when ${\cal C}_{12}\simeq 1$, and weakly connected when ${\cal C}_{12} \simeq 0$. 

A particular case of a strong connection between two points is the single-mode regime, for which the equality
${\cal C}_{12} = 1$ is satisfied, as we shall now see.
In a single-mode lossless cavity, the electric Green function can be expanded in the form 
\begin{align}
   \bG(\br_1,\br_2,\omega) =c^2 \, \be_M(\br_1) \be_M^*(\br_2)
   \times \left [ \mathrm{PV} \left (\frac{1}{\omega_M^2-\omega^2} \right ) \right.  \nonumber \\  \left.+ \frac{i\pi}{2\omega_n} \delta(\omega-\omega_M)\right ]
   \label{eq:G_modes}
\end{align}
where $\be_M(\br)$ is the eigenmode and $\omega_M$ the eigenfrequency of the vector Helmholtz equation~(see for example
Ref.~\cite{Carminati-review}).  Note that we use a tensor notation such that $ [\be_M(\br_1) \be_M^*(\br_2) ] \bv =
[\be_M^*(\br_2) \cdot \bv] \be_M(\br_1)$ for any vector $\bv$.  In the presence of weak losses (by absorption or
radiation), we can write
\begin{align}
   \bG(\br_1,\br_2,\omega) =c^2 \, \frac{\be_M(\br_1) \be_M^*(\br_2)}{\omega_M^2-\omega^2 - i\omega\gamma_M}
   \label{eq:G_single_mode}
\end{align}
where $\gamma_M$ is the mode damping rate. This phenomenological expression is built in such a way to recover
Eq.~(\ref{eq:G_modes}) in the limit $\gamma_M \to 0$, and is only valid for a large quality factor
$Q=\omega_M/\gamma_M$, and for near-resonant frequencies. A more general approach could be built using quasi-normal
modes~\cite{Sauvan2014}, but we assume here that the conditions of the phenomenological approach (large $Q$ regime) are satisfied.
Using the reciprocity theorem $\bG(\br_1,\br_2,\omega)=\bG^T(\br_2,\br_1,\omega)$ satisfied by the Green function, one
easily shows that $\mathrm{Im}[\be_M(\br_1) \be_M^*(\br_2)]=0$.  As a consequence, the projected CDOS $\rho_{12} =
(2\omega/\pi c^2)  \Im [ \bu_1 \cdot \bG (\br_1, \br_2, \omega) \cdot \bu_2]$, with $\bu_1$ and $\bu_2$ two real unit
vectors, reduces to
\begin{align}
   \rho_{12} = \frac{\gamma_M}{2\pi} \frac{\bu_1 \cdot [\be_M(\br_1) \be_M^*(\br_2)] \bu_2}{(\omega_M-\omega)^2 + \gamma_M^2/4}
\end{align}
where we have used $\omega \simeq \omega_M$ except in the resonant term. Using the definition of the tensor product,
this expression can be transformed into
\begin{align}
   \rho_{12} = \frac{\gamma_M}{2\pi} \frac{[\be_M^*(\br_2) \cdot \bu_2][\be_M(\br_1) \cdot \bu_1] }{(\omega_M-\omega)^2 + \gamma_M^2/4} .
\end{align}
Introducing the LDOS at positions $\br_1$ and $\br_2$
\begin{align}
  \rho_{11} = \frac{\gamma_M}{2\pi} \frac{|\bu_1 \cdot \be_M(\br_1)|^2}{(\omega_M-\omega)^2 + \gamma_M^2/4} \\
   \rho_{22} = \frac{\gamma_M}{2\pi} \frac{|\bu_2 \cdot \be_M(\br_2)|^2}{(\omega_M-\omega)^2 + \gamma_M^2/4} 
\end{align}
one immediately obtain
\begin{align}
   \rho_{12}^2 = \rho_{11} \rho_{22}
\end{align}
in the case of a single mode. Note that we have identified the square and the square modulus of the real-valued denominator in
$\rho_{12}$.
From the definition of the connectivity [Eq.~(\ref{def_connectivity})], we directly see that ${\cal
C}_{12}=1$ when the two points $\br_1$ and $\br_2$ are connected by a single weakly-dissipative mode.

\section{Connectivity and Anderson localization}

Using the connectivity as a measure of the mode connection between two points is of interest, for example,
to probe the existence of a single-mode regime, resulting from a specific design of a structure, or from a self-build
localization process. As an important application of the concept, we shall now show that the connectivity discriminates between diffusive
transport and Anderson localization in disordered media. Identifying an unambiguous marker of these two regimes remains
a challenging issue, in particular for electromagnetic waves. 

Before analyzing the relevance of the connectivity in this context, we give a qualitative picture of the difference
between diffusion and localization in terms of eigenmodes. In the diffusive regime, the eigenmodes overlap both in
frequency and space. At a given frequency, the eigenmodes are spatially extended, and any point in the medium is covered
by a large number of modes. Conversely, in the localized regime, at a given point and for a given frequency, no more
than one mode has a non-negligible contribution~\cite{Caze2013-sc}. 

To support this qualitative picture, we provide numerical simulations of scattering of electromagnetic waves in two dimensions (2D).
We restrict the simulations to 2D since the existence of Anderson localization of electromagnetic waves in three dimensions (3D)
remains an open question, while its existence in 2D has been proven~\cite{Laurent2007,Wiersma2011}. Although the method proposed
in this work could provide an unambiguous signature of localization even in 3D, our purpose here is to prove the principle
using numerical simulations in a 2D geometry for which the diffusive and localized regimes can be clearly identified. Indeed, even the
numerical proof of Anderson localization of electromagnetic waves in 3D remains a matter of debate~\cite{Skipetrov2014}, and 
we do not intend to solve this issue here.

\begin{figure}[htbp]
   \centering
   \includegraphics[width=\linewidth]{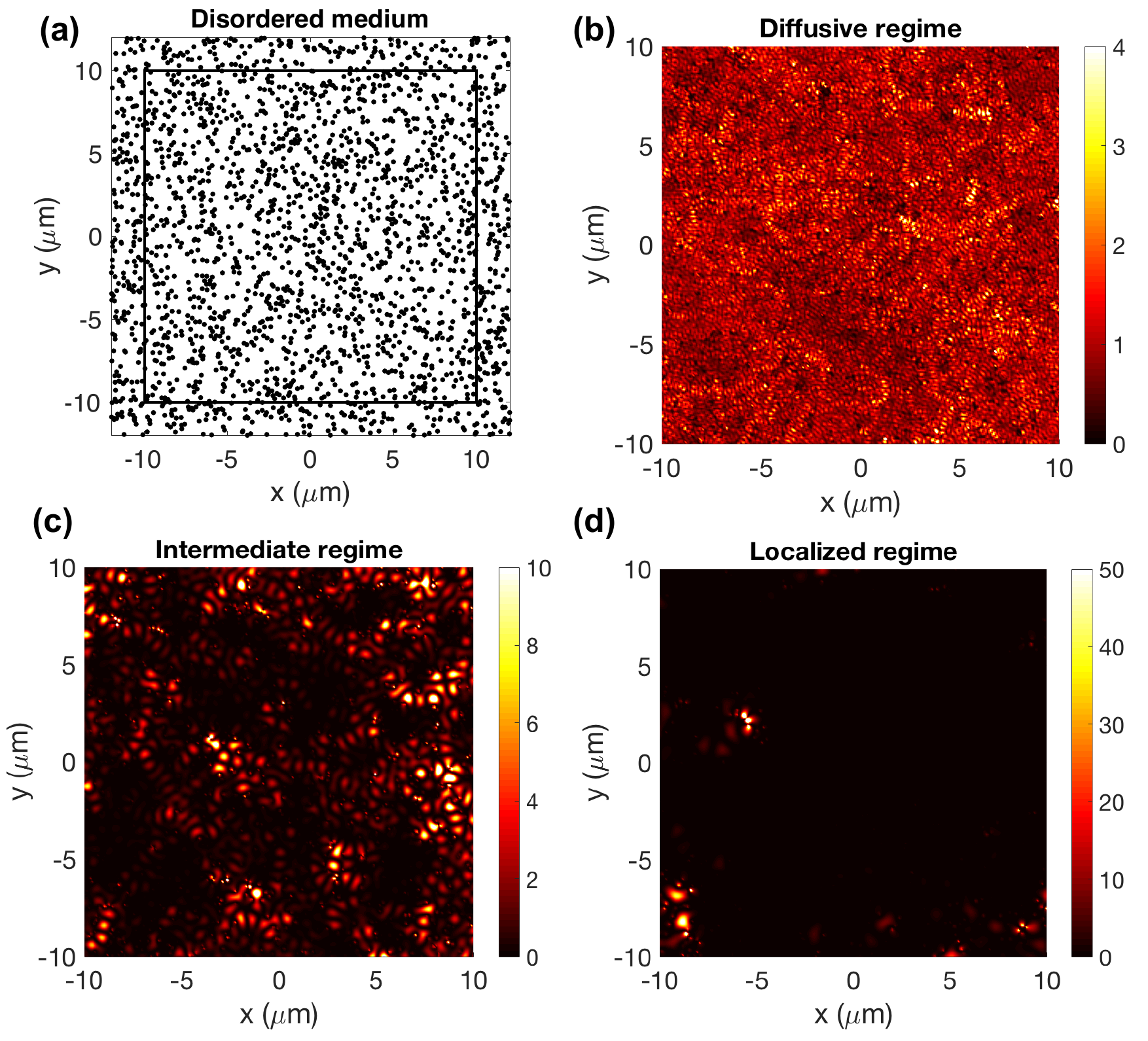}
   \caption{(Color online) (a)~Realization of a 2D disordered medium with $N=2292$ subwavelength scatterers. To avoid
   border effects, fields are calculated within the inner 20 $\mu$m by 20 $\mu$m square box defined by the solid line.
   (b-d)~Calculated LDOS maps. (b)~Diffusive regime ($\lambda = 400$ nm). (c)~Intermediate regime ($\lambda = 1000$ nm).
   (d)~Localized regime ($\lambda = 1500$ nm).}
   \label{LDOS_maps}
\end{figure}

For the numerical simulations, we consider a medium composed of randomly distributed non-absorbing
subwavelength scatterers. For illustrative purposes we consider TE polarized waves, with the electric field and the two
polarization directions $\bu_1$ and $\bu_2$ perpendicular to the plane containing the scatterers. The scatterers are
characterized by their electric polarizability $\alpha(\omega) = (2 \Gamma/k_0^2) (\omega_0 - \omega - i \Gamma/2)^{-1}$
with resonance frequency $\omega_0 = 3 \cdot 10^{15}$~s$^{-1}$, natural linewidth $\Gamma = 5 \cdot 10^{16}$~s$^{-1}$
and $k_0=\omega/ c=2\pi/\lambda$, where $\lambda$ is the wavelength in vacuum. This form of the polarizability is valid
for near-resonance frequencies and satisfies energy conservation (or equivalently the optical theorem). The surface
density of scatterers is $\rho = 3.98 \cdot 10^{12}$~m$^{-2}$, which corresponds to $N=2292$ point scatterers located in
a square domain with size $L=24$ $\mu$m [see Fig.~\ref{LDOS_maps}(a)]. From these parameters one can determine the
scattering mean free path $\ell_s = (\rho \sigma_s)^{-1}$, and estimate the localization length $\xi=\ell_s \exp(\pi k_0
\ell_s/2)$. For $\lambda=400$ nm, we get $\xi=4.16\cdot 10^{10} \mu$m $\gg L$ and the medium is in the diffusive regime,
while for $\lambda=1500$ nm, $\xi=1.5~\mu$m $\ll L$ and the medium is in the localized regime. The wavelength $\lambda
=1000$ nm provides an intermediate case for which $\xi = 19.6~\mu$m $\sim L$.  

We compute the field in the medium upon illumination by a single dipole source using the coupled dipole method, that has
been described in previous studies~\cite{Caze2013-sc}. From the field, one can deduce the Green function and the CDOS or
the LDOS based on Eq.~(\ref{def_CDOS}).  Maps of the LDOS inside the medium are displayed in Fig.~\ref{LDOS_maps}(b-d).
In order to avoid border effects and consider only bulk properties, we remove from the maps a 2 $\mu$m border [see
Fig.~\ref{LDOS_maps}(a)] that is larger than both $\ell_s = 993$ nm in the diffusive regime ($\lambda=400$ nm) and $\xi
=1.5~\mu$m in the localized regime ($\lambda = 1500$ nm). We clearly see that in the diffusive regime
[Fig.~\ref{LDOS_maps}(b)] the medium supports a spatially homogeneous distribution of LDOS, while in the localized
regime [Fig.~\ref{LDOS_maps}(d)] LDOS spots corresponding to localized modes are clearly visible, whereas a large part
of the sample is not covered by any eigenmode. 

In free space, for TE polarized waves in 2D, the connectivity for two points separated by a distance $r=|\br_1-\br_2|$
is simply ${\cal C}_0=\left| J_0(k_0 r) \right|$. In this simple case, ${\cal C}_0\simeq 1$ for positions separated by a
subwavelength distance ($k_0 r \ll 1$), while two positions far apart ($k_0 r \gg 1$) are poorly connected. From the
qualitative behavior described above, we can expect the connectivity ${\cal C}_{12}$ to change substantially when the
transport regime shifts from diffusive to localized. In the first case we await the connectivity to quickly decrease
with distance as in the vacuum case, due to the overlap of numerous eigenmodes. On the opposite, in the localized regime
where the modes are spatially separated, the connectivity is expected to fluctuate between ${\cal C}_{12} \simeq 1$ (for
two points in the same localized mode) and ${\cal C}_{12} \simeq 0$ (for unconnected points), the latter being more
likely for distance $|\br_1-\br_2| > \xi$. We have checked this behavior numerically by computing maps of ${\cal
C}_{12}$ versus $\br_2$, with $\br_1$ fixed at the center of the medium.

\begin{figure}[htbp]
   \centering
   \includegraphics[width=\linewidth]{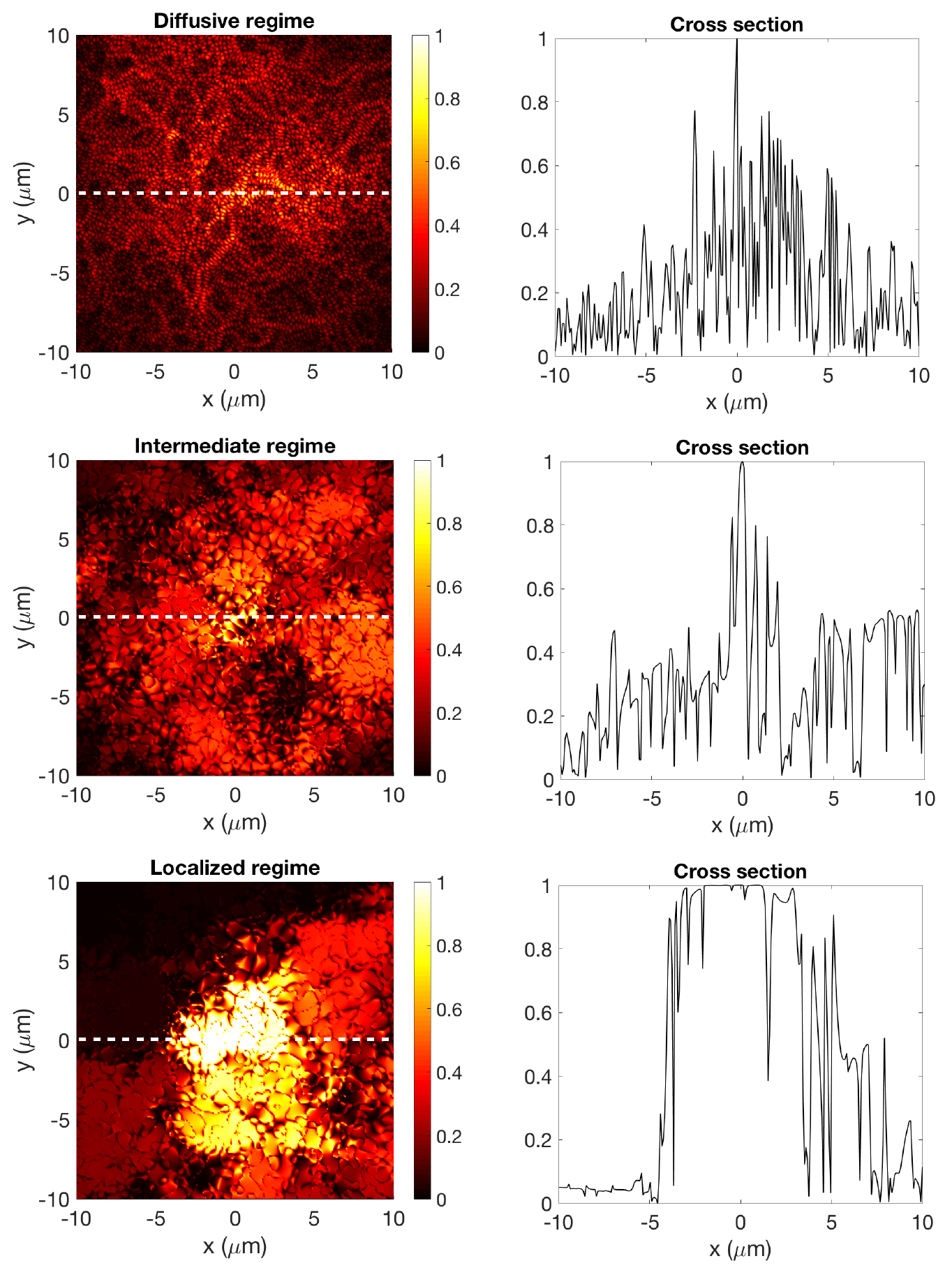}
   \caption{(Color online) Left column: Maps of the relative connectivity ${\cal C}_{12}$ versus $\br_2$, for a fixed
   position of $\br_1$ (chosen at the center), in a single realization of disorder. Right column: Cross sections along
   the line $y=0$. From top to bottom the system transits from the diffusive to the localized regimes, with the
   wavelength as a control parameter.}
   \label{ncdos_maps}
\end{figure}

The results are shown in Fig.~\ref{ncdos_maps}, with connectivity maps in the left column, and cross-sections along the
line $y=0$ in the right column. The top row correponds to the diffusive regime ($\lambda=400$ nm), the middle row to the
intermediate regime ($\lambda=1000$ nm), and the bottom row to the localized regime ($\lambda= 1500$ nm). The three maps
exhibit large qualitative differences: Values of ${\cal C}_{12}$ clearly below one are obtained in the diffusive regime
(top) except close to the origin, which corroborates the idea that any point in the medium is covered by a large amount
of modes. Conversely, in the localized regime (bottom) we observe that the relative connectivity saturates to its
maximum $({\cal C}_{12}=1)$ over distances $|\br_1-\br_2| \sim \xi$. This shows that the relative connectivity is a good
marker of the difference between the diffusive and the localized regimes, even in a single realization of the disordered
medium. This behavior has been systematically observed for many different realizations and results in different profiles
for the ensemble-averaged connectivity (see Appendix~\ref{ensemble-averaged}).

\section{Framework for practical implementation}

In this section, we shall discuss the relationship between the connectivity and observables in order to provide a
framework for practical implementations. Inspired by previous studies~\cite{Canaguier2016,Carminati2015}, a strategy
relies on placing two incoherent dipole sources inside the medium (or close to its surface), and measuring the time fluctuations 
in the total power emitted outside the medium. This strategy is here studied in the general case, that covers 2D and 3D geometries.
We also consider both classical emitters (e.g. two dipole antennas excited with mutually uncorrelated time-fluctuating currents) and 
quantum single-photon sources, thus providing tools for a practical implementation in different spectral ranges.

For classical emitters, fluctuations of the total emitted power are characterized by the second-order quantity
$G^{(2)}_\mathrm{class} = \overline{P^2} / \left( \overline{P} \right)^2$, where $P$ is the total power emitted by two
dipole sources, and $\overline{X}$ denotes the time average of $X$ over the temporal fluctuations of the sources. For
uncorrelated (incoherent) and similar sources, one shows that~\cite{Carminati2015}:
\begin{align}\label{G2class_g}
   G^{(2)}_\mathrm{class} (\br_1,\br_2,\omega) = 1 + \frac{1}{2}  \left ( 1-{\cal F}_{12}^2 \right ) {\cal C}_{12}^2
\end{align} 
where ${\cal F}_{12}$ is the LDOS contrast defined as
\begin{align}\label{contrast_def}
   {\cal F}_{12} =\frac{\left| \rho_{11} - \rho_{22}\right|}{\rho_{11} + \rho_{22}} .
\end{align}
Although this result can be directly deduced from the analysis in Ref.~\cite{Carminati2015}, it is derived in
Appendix~\ref{second_order_coherences} for the sake of consistency. The dimensionless quantity ${\cal F}_{12}$, that also lies
within $[0,1]$, measures the relative difference between the LDOS at positions $\br_1$ and $\br_2$. Values ${\cal
F}_{12} \simeq 0$ correspond to $\rho_{11} \sim \rho_{22}$, while ${\cal F}_{12} \simeq1$ when $\rho_{11} \ll \rho_{22}$
or $\rho_{11} \gg \rho_{22}$. From Eq.~(\ref{G2class_g}), one readily sees that the classical second-order function
$G^{(2)}_\mathrm{class}$ lies in $[1, 3/2]$ since both $(1-{\cal F}_{12}^2)$ and ${\cal C}_{12}^2$ are in $[0,1]$.

For two single-photon quantum emitters, the second order coherence function for measurements integrated over all output
channels is defined as $G^{(2)}_\mathrm{quant} = \langle P_2 \rangle / \langle P_1 \rangle^2$, where $P_1$ and $P_2$ are
the single and double photodetection operators integrated over all output channels, and the brackets denote quantum
expectation values. Denoting by $\bPhi_1(\br_1,\alpha_1)$ (resp. $\bPhi_2(\br_1,\alpha_1,\br_2,\alpha_2))$ the single
(resp. double) photodetection quantum operators, $\br_1$ and $\br_2$ denoting two detector positions and $\alpha_1$ and
$\alpha_2$ being two polarization states, one has $P_1=(\eps_0 c/2) \int_{S_1} \d \br_1  \sum_{\alpha_1} \bPhi_1
(\br_1,\alpha_1)$ and $P_2=  (\eps_0 c/2)^2 \int_{S_1} \d \br_1  \int_{S_2} \d \br_2 \sum_{\alpha_1,\alpha_2} \bPhi_2
(\br_1,\alpha_1,\br_2,\alpha_2)$. The summation over polarization, and the integration over two surfaces $S_1$ and $S_2$
enclosing the medium, define photodetection processes integrated over all output channels~\cite{Canaguier2016}. In the
case of two similar emitters initially in the excited state, the second order coherence function
$G^{(2)}_\mathrm{quant}$ can be expressed in terms of the LDOS contrast and the relative connectivity in the form
\begin{align}\label{G2quant_g}
   G^{(2)}_\mathrm{quant} (\br_1,\br_2,\omega) = \frac{1}{2}  \left (1-{\cal F}_{12}^2 \right ) \left ( 1+ {\cal C}_{12}^2 \right ) .
\end{align} 
This result is directly deduced from the analysis in Ref.~\cite{Canaguier2016}, and is also derived in
Appendix~\ref{second_order_coherences} for consistency.  Expression (\ref{G2quant_g}) shows that $0\leq
G^{(2)}_\mathrm{quant} \leq 1$, meaning an anti-bunching behavior in the emission from the two quantum sources. 

For both classical or quantum sources, we conclude that second-order coherence functions $G^{(2)}$ are fully determined
by the mode connectivity ${\cal C}_{12}$ and the LDOS contrast ${\cal F}_{12}$. From the maps of the LDOS presented
in Fig.~\ref{LDOS_maps}, we expect  ${\cal F}_{12}$ to increase when the system goes from the diffusive regime to the
localized regime. This is confirmed in the 2D numerical simulations, since the averaged value of ${\cal F}_{12}$ over the maps gives 0.29
for $\lambda=400$ nm (diffusion), 0.63 for $\lambda=1000$ nm (intermediate regime) and 0.74 for $\lambda=1500$ nm
(localization). As a consequence, since both quantities ${\cal C}_{12}$ and ${\cal F}_{12}$ have a very different
behavior in the diffusive and localized regimes, we expect $G^{(2)}$ to be quite a useful tool to discriminate between
the two regimes.

\begin{figure}[htb]
   \centering
   \includegraphics[width=\linewidth]{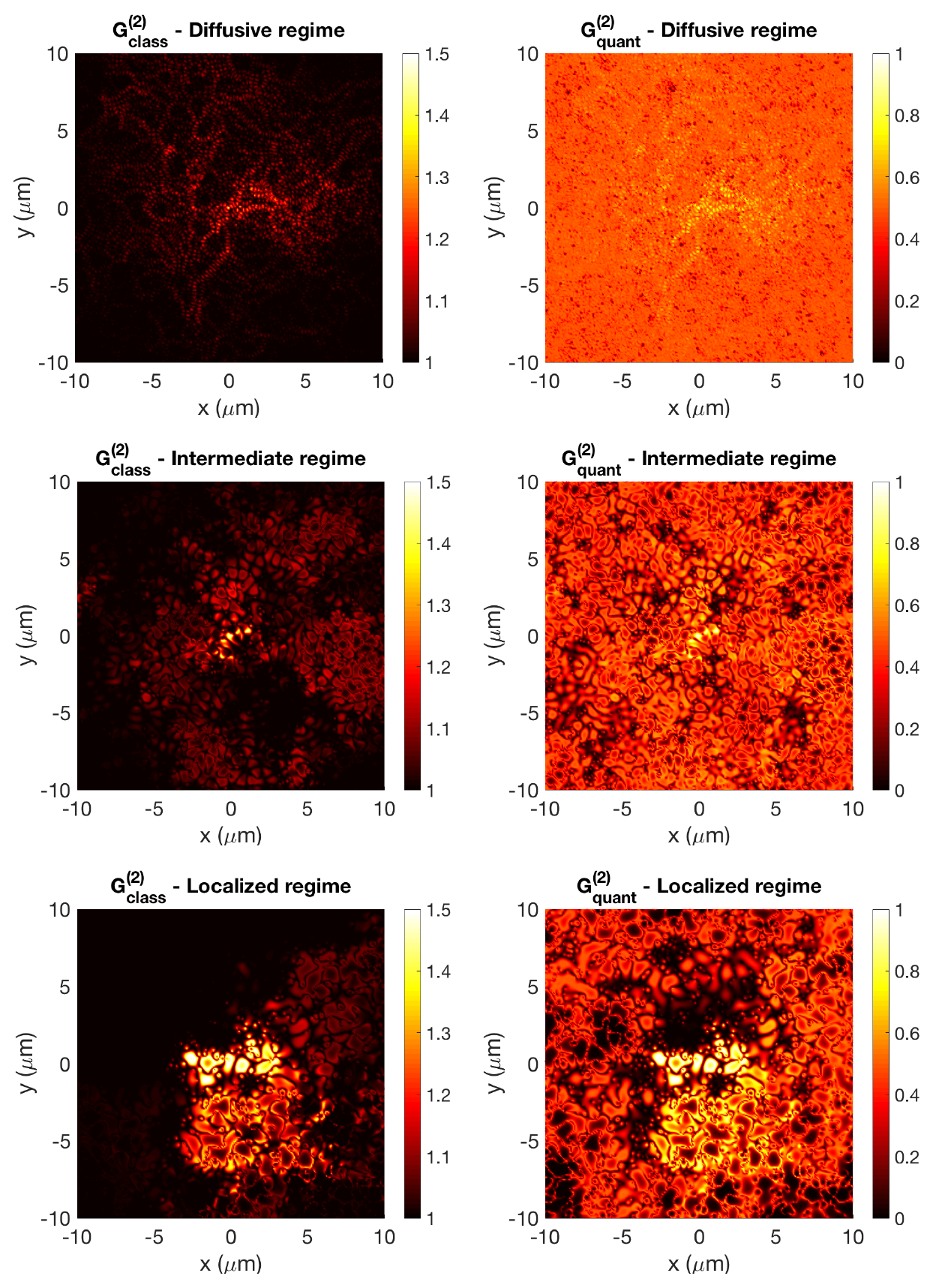}
   \caption{(Color online) Maps of the second-order coherence function $G^{(2)}_\mathrm{class/quant}
   (\br_1,\br_2,\omega)$ in a single realization of disorder, with $\br_1$ fixed at the center and $\br_2$ scanning the
   medium, with classical sources (left) and quantum sources (right).}
   \label{G2_maps}
\end{figure}

To confirm this assertion we present maps of the classical and quantum second-order coherence functions evaluated
 in one realization of disorder for a 2D system in
Fig.~\ref{G2_maps}, for the diffusive regime with $\lambda=400$ nm (top), the intermediate regime with $\lambda=$ 1000 nm
(middle) and the localized regime with $\lambda=1500$ nm (bottom). 

The maps for the classical second-order function $G^{(2)}_\mathrm{class}$ (left panels) are very similar to the maps of
the relative connectivity in Fig.~\ref{ncdos_maps}: Small values are obtained in the diffusive regime while large
clusters with high values of $G^{(2)}_\mathrm{class}$ appear in the localization regime. In the quantum case, the maps
of $G^{(2)}_\mathrm{quant}$ (right panels) are slightly different, as a result of the influence of two factors: The
relative connectivity ${\cal C}_{12}$ that governs the patterns close to the origin, corresponding to small distances
between the two observation points, and the LDOS contrast ${\cal F}_{12}$ that acts as a background for larger
separations. These differences can be traced back to the fact that the two second-order coherence functions include the
factor $(1-{\cal F}_{12}^2)$, but differ by their dependence on the relative connectivity. In particular, while the
variation of $G^{(2)}_\mathrm{class}$ is directly proportional to ${\cal C}_{12}^2$, the quantum second-order coherence
function, being proportional to $(1+{\cal C}_{12}^2)$, is less sensitive to ${\cal C}_{12}$. Moreover, for distant
observation points such that ${\cal C}_{12}\simeq 0$, one has $G^{(2)}_\mathrm{class} \rightarrow 1$, while
$G^{(2)}_\mathrm{quant} \rightarrow (1-{\cal F}_{12}^2)/2$, explaining why the background is affected by the spatial
distribution of the LDOS in the case of single-photon sources.

The similarity between the spatial bahavior of the classical coherence function $G^{(2)}_\mathrm{class}$ and the mode
connectivity suggests to use the measurement of $G^{(2)}_\mathrm{class}$ as a strategy to probe the appearance of
localized modes in disordered media, and in particular the transition from diffusive transport to Anderson localization.
The numerical study shows a robust signature of localization even in a single realization of disorder. As a supplemental
analysis, disordered-averaged profiles of $G^{(2)}_\mathrm{class}$ and $G^{(2)}_\mathrm{quant}$ versus the distance
$|\br_1 - \br_2 |$ between the observation points are presented in Appendix~\ref{ensemble-averaged}.

\section{Conclusion}

In summary, we have introduced the concept of mode connectivity as a measure of the connection between two points at a
given frequency. We have demonstrated the relevance of the connectivity in probing spatially localized modes in complex
media. In the case of electromagnetic waves, we have defined observables that directly depend on the connectivity, and
proposed schemes for practical implementations. The analysis could  allow one to define an unambiguous approach to probe
3D Anderson localization of electromagnetic waves. It can be extended to other kinds of waves, and is relevant to all
processes creating spatially localized modes.

\section*{ACKNOWLEDGMENTS}

This work was supported by LABEX WIFI (Laboratory of Excellence within the French Program ``Investments for the
Future'') under references ANR-10-LABX-24 and ANR-10-IDEX-0001-02 PSL*.

\appendix

\section{Derivation of the inequality between CDOS and LDOS}\label{cdos_inequality}

In this section we derive Eq.~(\ref{CS_inequality}). We consider two monochromatic electric dipole sources (frequency
$\omega$) with dipole moments $\bp_1=p_1 \bu_1$ and $\bp_2=p_2 \bu_2$, located at positions $\br_1,\br_2$ in a linear
and non-absorbing medium. The electric field radiated at point $\br$ can be written in terms of the Green function
\begin{align}
   \label{eq:field_Green}
   \bE(\br) = \mu_0 \omega^2 \bG(\br,\br_1,\omega) \bp_1 +  \mu_0 \omega^2 \bG(\br,\br_2,\omega) \bp_2 \ .
\end{align} 
The total power emitted by the two sources outside the medium is
\begin{align}\label{power_twosources}
   P &= \frac{\eps_0 c}{2} \int_S | \bE \ |^2 dS
\end{align}
where $S$ is a sphere with radius $R \to \infty$ that encloses the medium. Using Eq.~(\ref{eq:field_Green}), this can be rewritten as
\begin{align}\label{power_twosources_rho1}
   P &= \frac{\pi \omega^2}{4\eps_0} \left[ |p_1|^2 \rho_{11} + |p_2|^2 \rho_{22} + 2 \Re[p_1 p_2^*] \rho_{12} \right] 
\end{align}
where $\rho_{ij}=(2\omega/\pi c^2)  \Im [ \bu_i \cdot \bG (\br_i, \br_j, \omega) \cdot \bu_j] $ is the CDOS for $i\neq
j$ and the LDOS for $i=j$.  Assuming that the two dipole sources are in-phase ($p_2=\beta p_1$ with $\beta$ a real
number), we get 
\begin{align}
   P = \frac{\pi \omega^2}{4 \eps_0} |p_1|^2 \left[ \rho_{22}  \beta^2 + 2 \rho_{12}   \beta + \rho_{11} \right]
\end{align}
which is a second-order polynom in $\beta$. As the radiated power $P$ is positive, the determinant of this polynom must
be negative, which yields
\begin{align}
   (\rho_{12})^2 \leq \rho_{11} \rho_{22} .
\end{align}
This proves Eq.~(\ref{CS_inequality}).

\section{Second order coherences}\label{second_order_coherences}

In this appendix we recall the expression of the second-order coherence functions that were initially derived in
Refs.~\cite{Carminati2015} and~\cite{Canaguier2016} for classical and quantum sources, respectively.

\subsection{Classical sources}

We consider two incoherent classical sources located at positions $\br_1,\br_2$ in a linear and non-absorbing medium,
whose dipole moments $\bp_1, \bp_2$ are written as $\bp_k(t)= p_k e^{\imath \phi_k(t)} e^{-\imath \omega t} \bu_k$,
where $\phi_1(t), \phi_2(t)$ are slowly varying uncorrelated random phases and $\bu_1, \bu_2$ are fixed orientations.
Using Eqs.~(\ref{eq:field_Green}) and (\ref{power_twosources}) above, the total power emitted outside the medium reads
as
\begin{align}\label{power_twosources_rho}
   P &= \frac{\pi \omega^2}{4\eps_0} \left[ |p_1|^2 \rho_{11} + |p_2|^2 \rho_{22} + 2 \Re[p_1 p_2^* e^{\imath(\phi_1-\phi_2)}] \rho_{12} \right] 
\end{align}
where $\rho_{ij}=(2\omega/\pi c^2)  \Im [ \bu_i \cdot \bG (\br_i, \br_j, \omega) \cdot \bu_j] $ is the CDOS for $i\neq
j$ and the LDOS for $i=j$.  When averaging over time, $e^{\imath(\phi_1-\phi_2)}$ vanishes in the interference term
which leads to
\begin{align*}
   \overline{P} = \frac{\pi \omega^2}{4\eps_0} \left[ |p_1|^2 \rho_{11} + |p_2|^2 \rho_{22} \right] ~ .
\end{align*}

On the opposite, when looking at the variance of $P$, a cross-product of interference terms survive the averaging
process and
\begin{align*}
   \overline{P^2}- \left( \overline{P} \right)^2= 2 \left(\frac{\pi \omega^2}{4\eps_0}\right)^2 |p_1 p_2|^2 (\rho_{12})^2   
\end{align*}
which leads to the second order coherence of the total emitted power
\begin{align*}
   G^{(2)}_\mathrm{class} &= \frac{\overline{P^2}}{ \left( \overline{P} \right)^2} =1 + \frac{ 2 |p_1 p_2|^2 (\rho_{12})^2 }{\left[ |p_1|^2 \rho_{11} + |p_2|^2 \rho_{22} \right]^2} ~ .
\end{align*}
For emitters with similar amplitude ($|p_1|=|p_2|$), it simplifies to 
\begin{align}\label{result_G2_class}
   G^{(2)}_\mathrm{class} &= 1+ \frac{1}{2} \left[ 1- {\cal F}_{12}^2 \right] {\cal C}_{12}^2
\end{align}
where
\begin{align*}
   {\cal F}_{12} = \frac{| \rho_{11} - \rho_{22} |}{\rho_{11} + \rho_{22}} ~ ~  \mathrm{and} ~ ~ {\cal C}_{12}= \frac{| \rho_{12} |}{\sqrt{\rho_{11}  \rho_{22}}}
\end{align*}
are the LDOS contrast and the mode connectivity between locations $\br_1$ and $\br_2$, respectively. The advantage of
expression (\ref{result_G2_class}) is that it highlights that for the coherence to reach high values
($G^{(2)}_\mathrm{class} \simeq 3/2$), the emitters must be located at positions that are well-connected (${\cal
C}_{12}\simeq 1$) and where the LDOS is well-balanced (${\cal F}_{12} \simeq0$).

\subsection{Quantum sources}

A similar derivation can be conducted for the case of quantum emitters, in which case the positive-frequency component
of the electric field operator can be connected to the source operators using the same Green tensors (see
Ref.~\cite{Canaguier2016} for more details):
\begin{align*}
   \bE^{(+)} (\br) = \mu_0 \omega^2  \sigma_1^- \bG (\br, \br_1) \bp_1 + \mu_0 \omega^2  \sigma_2^- \bG(\br,\br_2) \bp_2 .
\end{align*} 
Then the photodetection of one photon at position $\br_a$, with polarization state $\alpha_a$ along $\be_a$, is
described by the operator $\bPhi_1(\br_a, \alpha_a) = E_a^\dagger E_a$, where $E_a=\be_a \cdot \bE^{(+)}(\br_a)$.
Similarly, one defines the photodetection of two photons at positions $(\br_a, \br_b)$, with respective polarizations
$(\alpha_a,\alpha_b)$, with the operator $\bPhi_2 (\br_a,\alpha_a,\br_b,\alpha_b) =E_a^\dagger E_b^\dagger E_b E_a$.
When integrating over all possible directions and polarizations one gets the operators
\begin{align*}
   P_1&=\frac{\eps_0 c}{2} \int_{S_a} \d \br_a  \sum_{\alpha_a} \bPhi_1 (\br_a,\alpha_a)\\
   P_2&=  \left( \frac{\eps_0 c}{2} \right)^2 \int_{S_a} \d \br_a  \int_{S_b} \d \br_b \sum_{\alpha_a,\alpha_b} \bPhi_2 (\br_a,\alpha_a,\br_b,\alpha_b) 
\end{align*}
where the prefactors are chosen to define observables corresponding to radiated power.  Assuming the two emitters in the
excited state, the probabilities to detect one or two photons over all output channels can then be simply expressed from
the LDOS and CDOS at the positions of the emitters:
\begin{align*}
   \langle P_1 \rangle &=  \frac{\pi \omega^2}{4\eps_0} \left( |p_1|^2 \rho_{11} + |p_2|^2 \rho_{22} \right) \\
   \langle P_2 \rangle &=  2 \left(\frac{\pi \omega^2}{4\eps_0} \right)^2  |p_1 p_2|^2 \left [\rho_{11}\rho_{22} + (\rho_{12})^2 \right] ~ .
\end{align*}
While the first quantity is similar to $\overline{P}$ for classical emitters, the second quantity differs as it does not
contain terms with $(\rho_{ii})^2$, due to the fact that the sources are single-photon emitters. The second-order
coherence for an emission integrated over directions and polarization is then 
\begin{align*}
   G^{(2)}_\mathrm{quant} &= \frac{\langle P_2 \rangle}{ \langle P_1 \rangle^2}  = \frac{ 2 |p_1 p_2|^2 \left[ \rho_{11} \rho_{22} + (\rho_{12})^2 \right]}{\left[ |p_1|^2 \rho_{11} + |p_2|^2 \rho_{22} \right]^2}
\end{align*}
which for emitters with similar amplitudes simplifies to 
\begin{align}\label{result_G2_quant}
   G^{(2)}_\mathrm{quant} = \frac{1}{2} \left[ 1- {\cal F}_{12}^2 \right]\left[ 1+  {\cal C}_{12}^2 \right] ~ .
\end{align}
Again, the later expression is quite useful as it readily shows that to the second-order coherence reaches its maximum
values ($G^{(2)}_\mathrm{quant} \simeq 1$) when ${\cal F}_{12}\simeq 0$ and ${\cal C}_{12} \simeq 1$.

\section{Ensemble-averaged quantities}\label{ensemble-averaged}

In order to overcome the dependence of the results on a particular realization of the scatterers positions, we have
repeated the process of generating  the maps of the various observables over 240 realizations of the disordered medium.
For an observable $X(\br_1,\br_2)$, we first compute an averaged of this ensemble for each positions to produce a single
averaged map. The points are then grouped by concentric rings to get a profile $\langle X \rangle(r)$ as a function of
the distance $r=|\br_1-\br_2|$ between the two positions. This enables to check that evaluations can rely in practice on
a single realization of disorder, and to observe the qualitative evolution of an observable $X(\br_1,\br_2)$ when the
system goes from diffusive to localized regime beyond the  particular realization of disorder.

\subsection{Mode-connectivity}

We first consider the mode connectivity ${\cal C}_{12}$ which is the absolute value of the CDOS between two positions
normalized by their LDOSs and takes values between 0 and 1. Maps of this quantity for a single realization of disorder
are presented in the main text, and shows a qualitative change when going from the diffusive to the localized regime: In
the diffusive regime ${\cal C}_{12}$ takes values clearly below one except close to the origin, while in the localized
regime this quantity saturates for distances larger than the localization length $\xi$. 

\begin{figure}[htbp]
   \centering
   \includegraphics[width=\linewidth]{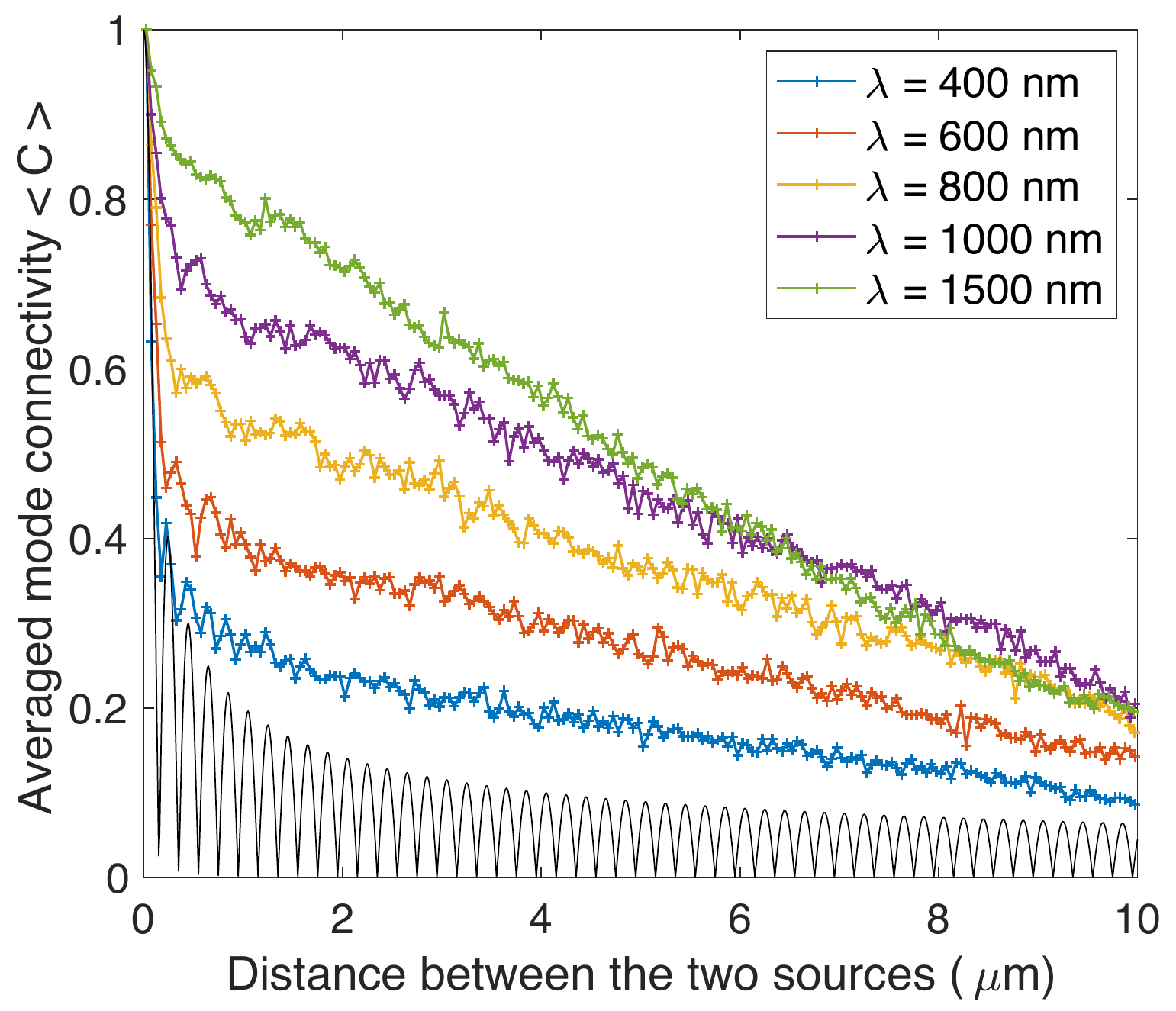}
   \caption{Estimate of the averaged mode connectivity profile $\langle {{\cal C}_{12}} \rangle (r)$ over 240
   realizations of disorder, as a function of the distance $r$ between the two locations, for several values of the
   wavelength $\lambda$. The thin solid line at the bottom is $|J_0(k_0 r)|$ for $\lambda=400$ nm, which recreates the
   connectivity of vacuum, for comparison of the behavior at small distances.}
   \label{ensemble_C}
\end{figure}

The averaged profile $\langle {\cal C}_{12} \rangle(r)$ is presented in Fig.~\ref{ensemble_C}.

In the diffusive regime, $\langle {{\cal C}_{12}} \rangle(r)$ quickly decreases with some oscillations, a reminder of
the vacuum case where $\mathcal{C}_0 (r)=|J_0(k_0r)|$ for 2D TE modes. When the system enters the localization regime
the drop becomes weaker as expected and high values are obtained  up to several micrometers of distance.

\subsection{Classical coherence}

We do the same treatment for the second-order coherence $G^{(2)}_\mathrm{class}$ in the case of two classical emitters.

\begin{figure}[htbp]
   \centering
   \includegraphics[width=\linewidth]{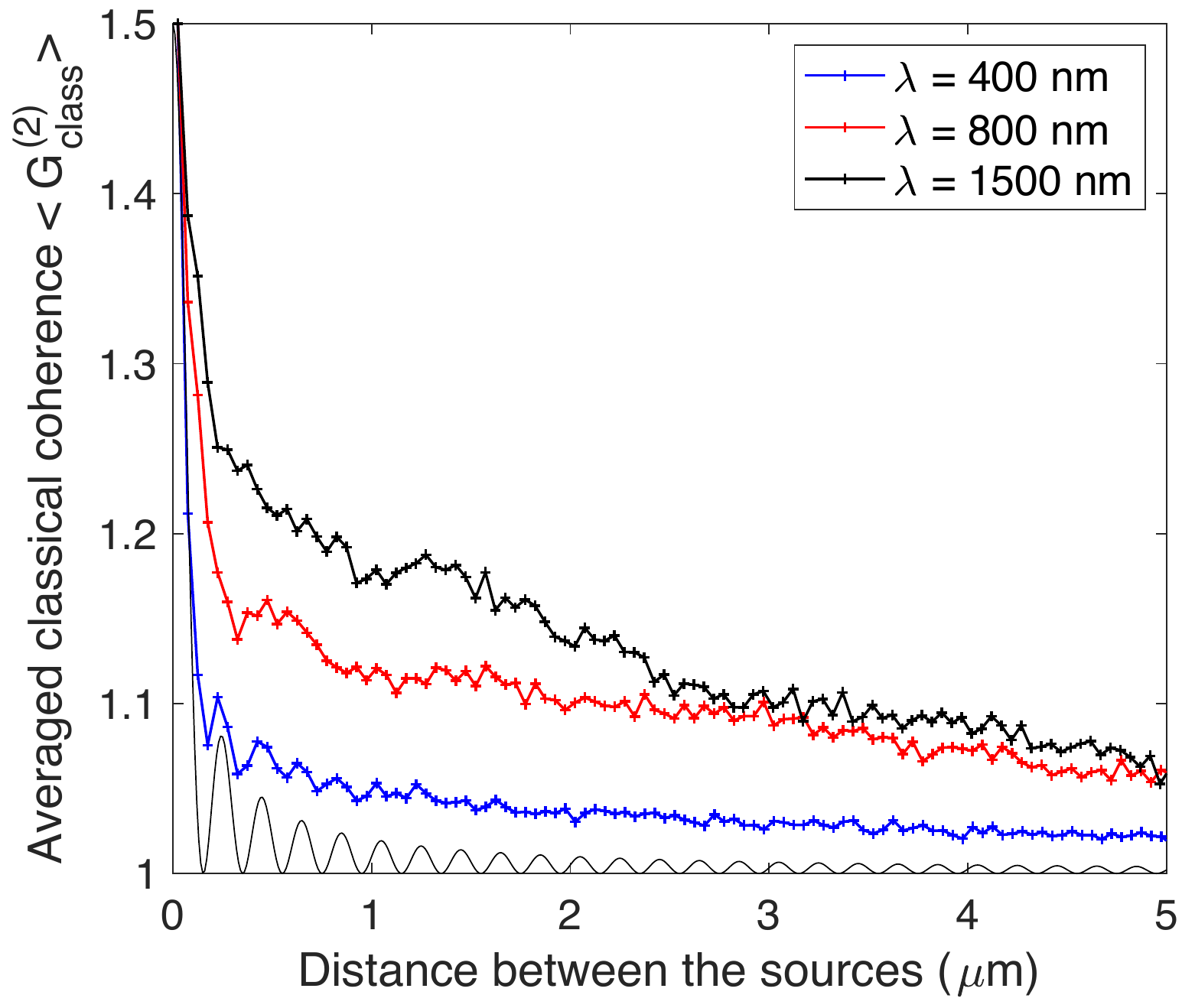}
   \caption{Estimate of the averaged second-order coherence for classical sources $\langle G^{(2)}_\mathrm{class}
   \rangle(r)$ as a function of the distance $r$ between the two sources, for several values of the wavelength
   $\lambda$.  The thin solid line at the bottom is $1+|J_0(k_0 r)|^2/2$ for $\lambda=400$ nm, which recreates the case
   of vacuum, for comparison of the behavior at small distances.}
   \label{ensemble_G2class}
\end{figure}

The averaged profile $\langle G^{(2)}_\mathrm{class} \rangle (r)$ is presented in Fig.~\ref{ensemble_G2class} and
corroborates the observation made for a single realization of disorder: Due to the large clusters of high values
appearing in the localization regime, the profile increases with the wavelength. In the diffusive regime the averaged
second-order coherence decreases strongly with small oscillations that are also present in the case of vacuum due to the
varying mode connectivity. 

\subsection{Quantum coherence}

\begin{figure}[h!]
   \centering
   \includegraphics[width=\linewidth]{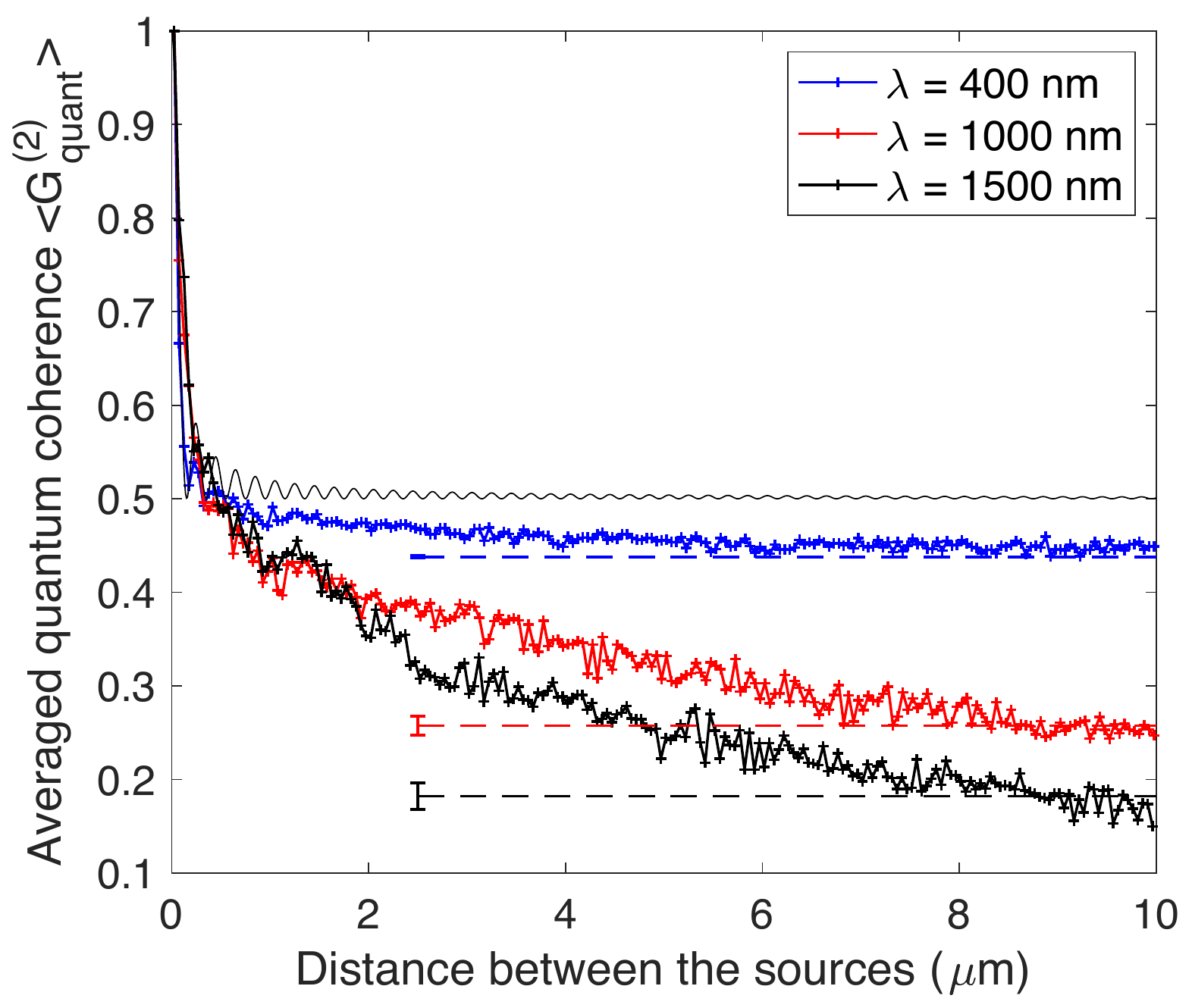}
   \caption{Estimate of the averaged second-order coherence for quantum sources $\langle G^{(2)}_\mathrm{quant} \rangle
   (r)$ as a function of the distance $r$ between the two sources, for several values of the wavelength $\lambda$.
   Dashed-lines represent estimates of the averaged value of $\langle (1-{{\cal F}_{12}^2})/2 \rangle$ and the error
   bars show the standard deviation when using one realization of disorder only.  The thin solid line is $(1+|J_0(k_0
   r)|^2)/2$ for $\lambda=400$ nm, which recreates the case of vacuum.
   \label{ensemble_G2quant}}
\end{figure}

The same analysis can be conducted for the case of quantum emitters. The profile of the averaged second-order coherence
$\langle G^{(2)}_\mathrm{quant} \rangle (r)$ is presented in Fig.~\ref{ensemble_G2quant} as a function of the distance
between the sources. 

This time, while for small distances there is no clear difference in the averaged coherence $\langle
G^{(2)}_\mathrm{quant} \rangle$ between the diffusive and localized regimes, for large distances one has ${{\cal
C}_{12}} \rightarrow 0$ and the second-order coherence heads towards $(1-{{\cal F}_{12}^2})/2$ which decreases from 0.5
for vacuum to lower values when the system goes from the diffusive to the localized regime. This asymptotic behavior is
confirmed by numerical estimation of $\langle (1-{{\cal F}_{12}^2})/2 \rangle$ presented in dashed horizontal lines of
corresponding wavelengths. 


\end{document}